# Unexpected Surface Implanted Layer in Static Random Access Memory Devices Observed by Microwave Impedance Microscope

W Kundhikanjana[1], Y Yang[1], Q Tang[2], K Zhang[2] K Lai[1], Y Ma[1]
M A Kelly[1], X X Li[2] and Z-X Shen[1]

[1]Department of Applied Physics and Geballe Laboratory for Advanced Materials, Stanford University, Stanford, California 94305, USA

[2]State Key Lab of Transducer Technology, Shanghai Institute of Microsystem

and Information Technology, Chinese Academy of Sciences, Shanghai 200050 China

Email: wkundhik@stanford.edu; zxshen@stanford.edu

**Abstract.** Real-space mapping of doping concentration in semiconductor devices is of great importance for the microelectronic industry. In this work, a scanning microwave impedance microscope (MIM) is employed to resolve the local conductivity distribution of a static random access memory (SRAM) sample. The MIM electronics can also be adjusted to the scanning capacitance microscopy (SCM) mode, allowing both measurements on the same region. Interestingly, while the conventional SCM images match the nominal device structure, the MIM results display certain unexpected features, which originate from a thin layer of the dopant ions penetrating through the protective layers during the heavy implantation steps.

1. Introduction

The ever-shrinking feature size of state-of-the-art semiconductor devices demands local probes with nanoscale resolution for design and characterization purposes. To date, two widely deployed electrical scanning probe techniques are scanning spread resistance microscope (SSRM) [1,2] and scanning capacitance microscopy (SCM)[1–3], both built on top of the atomic force microscope (AFM) platform. In SSRM, the high resolution conductivity map is measured through the DC current flow from the probe tip to a remote contact electrode. Such experiments are difficult for samples with native oxides. And the results on complex device structures are hard to interpret due to the complicated current path. SCM, on the other hand, overcomes the limitation by using a high-frequency (often at the microwave range) capacitance sensor to obtain the local electrical information. Because of this improvement, SCM has found several applications in characterizing deposited oxides, determining carrier types and concentrations, and performing failure analysis[1–3]. However, most commercial SCMs do not directly measure the tip-sample capacitance because of the strong background from unshielded probes, but rather the capacitance change induced by a low-frequency modulation (around 10 kHz) that drives the semiconductor underneath into accumulation or depletion regimes. Such a differential capacitance, or dC/dV, measurement results in non-monotonic response as a function of doping concentration, since low dC/dV signals are expected for both metallic (e.g., heavily doped) and insulating (e.g., dielectrics or depletion) regions. Furthermore, the large AC voltages may modulate a thick layer (up to micrometers) of the underlying carrier densities and complicate the data interpretation. As a result, SCM is still regarded as a qualitative local tool in the semiconductor industry[2].

The recently introduced scanning microwave impedance microscopy (MIM) further eliminates the above problems by directly measuring the real (MIM-Re) and imaginary (MIM-Im) components of the tip-sample impedance[4]. The technique is based on a near-field interaction at 1 GHz frequency, in which spatial resolution is determined by the tip apex size, currently about 50 nm. A notable innovation here is our shielded cantilever design, which greatly reduces parasitic tip-sample coupling that leads to topographical artifacts and environmental noise pickup[5]. With this feature, it is now straightforward to detect tiny tip-sample capacitance change – as small as 1aF[4]. The elimination of low-f modulation ensures the monotonic response in the MIM-Im channel as a function of the local conductivity. Finally, when necessary, the dC/dV experiment is easily performed in MIM for additional information, such as carrier type, of the sample.

In this paper, we demonstrate the MIM experiments on a static random access memory (SRAM) sample. Using two standard staircase samples, we confirm that the MIM-Im response is monotonic as a function of carrier densities, while the dC/dV amplitude peaks at an intermediate doping level. The MIM images on the SRAM devices show features unexpected from the nominal doping maps, which are attributed to a thin layer of dopants that went through protective layers during the ion-implantation processes. We explain why such surface effects are not captured by the dC/dV measurement.

2. Experimental Setup and Standard Samples

The schematic of our MIM setup is shown in Fig. 1a. Details of the electronics[4] and our shielded cantilever probes[5] can be found elsewhere. Microwave signal at 1 GHz is sent into a shielded probe and measure a reflected signal. The low input power of -14 dBm was used. As previously discussed in Ref [6], the MIM-Im signal, which is proportional to the tip-sample capacitance, is a monotonic function of the sample conductivity, while the MIM-Re signal, which reflects the resistive loss, peaks at intermediate doping levels. Only MIM-Im images are reported in the rest of the paper for a clear comparison with the dC/dV data. Note that the latter can be easily achieved by applying an AC modulation (typical $V_{AC}$ = 0.5 V at 5 kHz) on the bias-T and demodulating the MIM-Im signal with a lock-in amplifier.

The difference between MIM and dC/dV measurements is illustrated in Fig. 1 using standard p-type and n-type epi-layer staircase silicon wafers (IMEC, Belgium, part number T8_3 and ST3). The samples are mounted on the side with silver epoxy for good electrical grounding, and the side surfaces are polished before scanning. Fig. 1b (1f) shows MIM images of the p-type (n-type) sample. Each layer is labeled with only the order of magnitude of the doping concentration, with the exact numbers and corresponding conductivities listed in the caption. For comparison, dC/dV images on both samples are shown in Fig. 1c and Fig. 1g, with typical line cuts plotted in Fig. 1d and 1h.

To further emphasize the monotonic response to local conductive of MIM. We plotted average MIM (dC/dV) signal as a function of local conductivity in Fig. 1e (1i). Simulated MIM signal using finite element analysis is also plotted in Fig. 1e [4]. Both the simulation and the experimental data show that the MIM signal increases monotonically as increasing conductivity and doping concentration. However, deviation from the simulated response is observed at conductivity ~ 10 S/cm, which is likely due to carrier redistribution under the influence of the tip. More rigorous calculation is required to fully match the experimental signal. Meanwhile, in Fig. 1i, the dC/dV amplitude reaches a maximum around 50 S/cm ($10^{18}$ $cm^{-3}$) for both carrier types. The non-monotonic dC/dV response is easily understood because neither heavily-doped nor nearly-intrinsic semiconductors can have appreciable capacitance changes induced by the AC modulation. Monotonic response to the local conductivity is a clear advantage of MIM over the dC/dV measurement. Moreover, a flat-band DC voltage on top of the AC modulation is often needed in conventional SCM, which can be affected by the oxide thickness, the work function of the tip, and the carrier type[7,8]. In the next section, we will also show that, due to the strong modulation of carrier densities underneath the tip, the SCM may miss some surface effects in complicated structures, which are readily captured by our MIM.

3. Results and Discussions

Fig. 2a and 2b show the simultaneously taken AFM and MIM images of a SRAM sample provided by Bruker Corporation, California. Similar samples have been thoroughly studied by many techniques[9–11] and are currently used as test samples for commercial

SCMs (see for example Ref [12]), and thus is a good standard to test the relative merit of MIM. The dielectric layer and polysilicon gate are completely removed, allowing access to the underlying silicon layer. A cartoon of the designed sample structure is sketched in Fig. 2c, showing two types of MOSFETs, hereafter labeled as rectangular (NMOS) and H-shaped (PMOS) devices. Table 1 lists the carrier types and doping levels of individual regions[10]. As discussed below, these sample parameters match the dC/dV images taken by both our electronics and commercial SCMs (not shown). The nominal conductivities derived from Table 1 also explain most of the MIM results. The n+ and p+ electrodes, with rough surfaces due to the implantation damages, are brightest in Fig. 2b because of the high conductivity. Note that the surface roughness, which affects the tip-sample contact areas in the scanning, does couple into the MIM images. The signal decrease in the narrow lightly doped drain (LDD) and channel regions, both resolved in Fig. 2b. In between the p-epi and n-well areas, a low conductivity depletion region is observed as a vertical thick dark line.

Two surprising features, however, are identified in Fig. 2b. First, except for the two vertical lines inside the rectangle, bright regions around the source and drain electrodes are seen for the rectangular devices. This is in direct contradiction to the map in Fig. 2c, and was not reported by previous studies[9,10]. The second effect is relatively subtle. According to Table 1, the n and p channels and the n-well should have similar conductivities, which are higher than that of the p-epi layer. Our data, on the other hand, show systematically lower signals in the channels than in the p-epi or n-well background. We emphasize that the dark feature in the rectangle device and the channels are not due to a dead layer at the surface because increases in the conductivity can be observed by imaging with a fixed bias on the tip ($V_{tip}$). Fig. 2d shows examples of the MIM images with $V_{tip}$ = -3 V (left) and -1 V (right). The entire p-epi region becomes brighter as well as the channel regions. Negative voltage induces charge accumulation at the p-epi and the n-channels regions for the rectangular devices; thus higher contrast is observed. For the H-shaped device, lower conductivity is observed for the n-well region.

In order to understand these two peculiar features, one needs to make a clear distinction between the MIM and dC/dV measurements. In conventional SCM experiments, the DC offset and the low-f AC modulation, both in the order of volts, drive the charge carriers in and out of the semiconductor underneath the tip. The MIM probe, on the contrary, only applies millivolts of microwave excitation to the sample surface. As a result, MIM is more sensitive to a thin surface layer, while SCM detects the non-linear effect on a thicker layer modulated by the bias voltage. It is therefore reasonable to associate the observed anomalies in MIM with effects very close to the surface. In addition, the double line features are sharper compared with other structures and the alignment with the p-channels, suggesting that the relevant step had to occur after the gate structure was formed. The only possible processes are the implantation of the electrodes, which also involved high-energy implantation.

In Fig. 3, we illustrate the cross section (a) and the top view (b) of the sample during the heavy ion-implantation steps that create the electrodes. The n-type and the p-type implants were two separate steps, with photoresist protecting the opposite MOSFETs. During these procedures, the electrodes were covered with a thin oxide layer; the channels had both the oxide and poly-silicon gate on the top; and the rest of the device were protected with thick field oxide (Fig. 3a). Ideally, the implanted ions can only go through the thin oxide and should be blocked by the thick field oxide and the poly-gate. In the actual fabrication, however, the field oxide could be thinner than the desired value, for example from over polishing, leading to penetration of a small amount of high-energy ions through the protective layers. In the n or p-channels, these leak-through dopants compensate the channel carriers, resulting in lower conductivity on the very surface. For the rectangular device, the same effect also occurs underneath the field oxide, which is responsible for the higher conductivity seen in the MIM data. The double vertical lines inside the rectangle were protected by both the field oxide and the poly-silicon gate; therefore the conductivity here was not altered. Such surface implanted layers are not seen around the H-shaped devices because of the shallower profile for the p+ implants. Fig. 3c shows simulated doping profile for n+ and p+ implants (TSUPREM3, Synopsys, Inc.) in $SiO_2$ using parameters from Ref. [10], which were 4 x $10^{15}$ $cm^{-2}$ $As^+$ at 100 keV for the n+ regions and 2 x $10^{15}$ cm$^{-2}$ BF$_2^+$ 45 keV for the p+ regions, and a typical annealing condition of 900$^o$C and 30 mins [13].

As a final remark, the dC/dV amplitude and phase images taken by the same tip are shown in Fig, 4a and 4b, which are comparable to results from commercial SCMs[12]. For easy comparison with Fig. 2b, the color scale in Fig. 4a is chosen such that regions with lower signals appear brighter. It is obvious that the dC/dV response agrees with the nominal doping level in Table 1. The moderately doped p-epi and n-well regions show higher dC/dV amplitudes than the depletion region in between them and the heavily doped n+ and p+ electrodes. The phase image provides additional information about the carrier types, with opposite signals between the p-epi and n-well and signals on the electrodes in the middle. We emphasize that the bright areas around the n-electrodes and the double dark lines inside the rectangle are not evident in Fig. 4. And the n and p channels show the same dC/dV responses as the surrounding wells, rather than lower signals in the MIM images. We can then conclude that the thin surface implanted layers cannot be captured by the SCM mode, which detects the capacitance change within a relatively thick layer under the tip. With ever shrinking device dimension, such sensitivity to ultra-thin layers near the surface and its combination with the SCM data, which can be recorded in conjunction with MIM, would add important insights to the understanding of device structures.

4. Conclusions

We have shown the microwave imaging on staircase and SRAM samples in both the linear impedance and dC/dV modes. The additional features observed in the SRAM devices are associated with surface implanted dopants, which have not been reported before by other scanning probes. Although the size of the devices in this work is large compared to today's state of the art, it is a standard to compare existing techniques. We also stress that there is no fundamental limit for a near-field based technique such as MIM to reach 10 nm spatial resolution. Moreover, the sensitivity to surface conductivity may prove critical for failure analysis. With both capabilities of measuring the surface impedance by MIM and carrier types by SCM, our technique clearly demonstrates great potentials for applications in the semiconductor industry.

Acknowledgement

The authors would like to thank Dr. Craig Nakakura for helping us understand the structure of SRAM devices and Dr. Chamin Su from Bruker Corporation, California, for sharing the n-type staircase and the SRAM samples. This work is supported by Center of Probing the Nanoscale PHY-0425897 and NSF grants DMR-0906027.

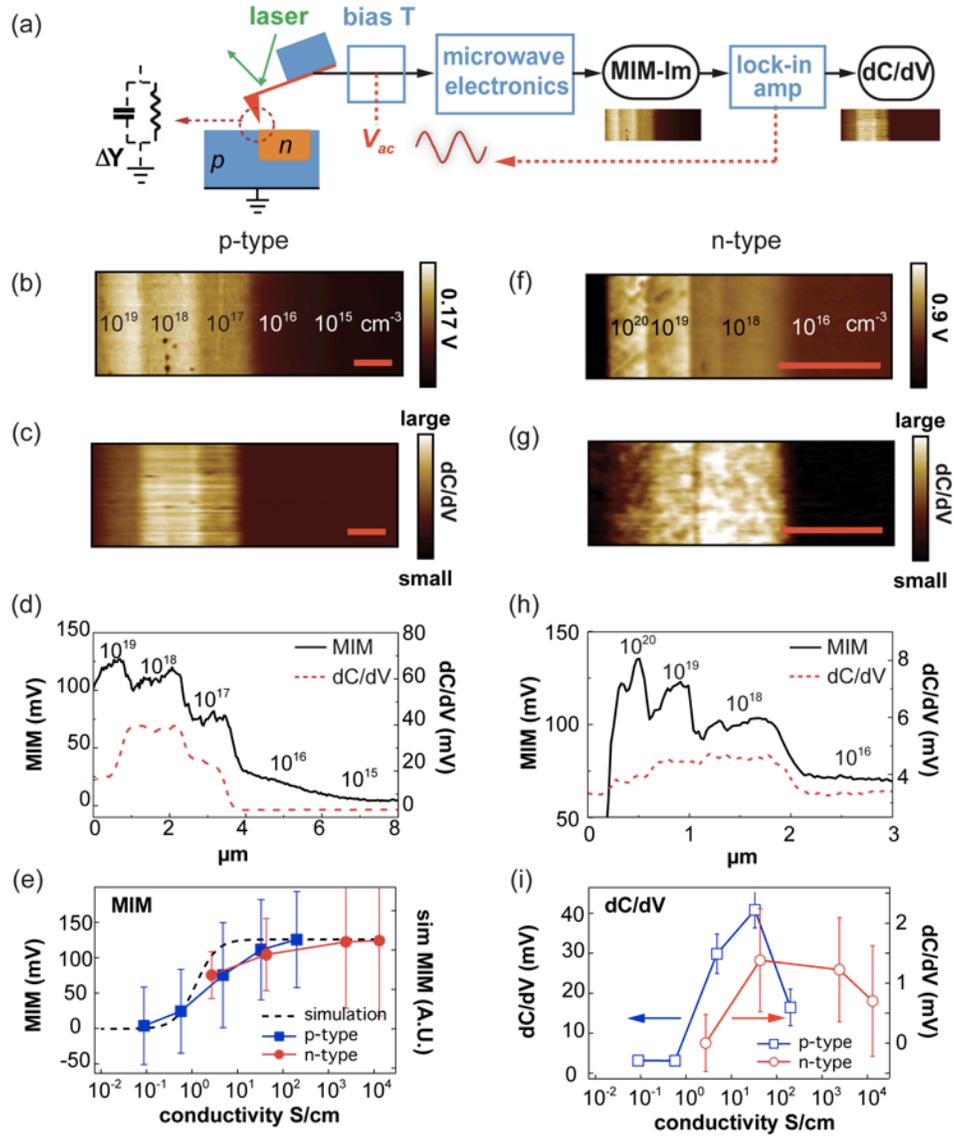

Figure 1. (Color online) (a) Schematic of MIM and dC/dV measurement. (b – d) MIM, dC/dV images, and single line cuts (MIM in solid line and dC/dV in dashed line) of the p-type IMEC staircase sample. (f – h) Same for the n-type IMEC sample. The hole concentrations in (b) (and conductivity, $\sigma$) from left to right are $2 \times 10^{19}$ ($2 \times 10^2$ S/cm), $1.8 \times 10^{18}$ (33 S/cm), $1.0 \times 10^{17}$ (4.8 S/cm), and $7.9 \times 10^{15}$ (0.57 S/cm), respectively. The electron densities (and $\sigma$) from left to right are $1 \times 10^{20}$ ($1.3 \times 10^3$ S/cm), $1.5 \times 10^{19}$ ($2.4 \times 10^2$ S/cm), $2.0 \times 10^{18}$ (60 S/cm), $1.0 \times 10^{18}$ (43 S/cm) and $1.5 \times 10^{16}$ (2.7 S/cm). (e) Average MIM signal as a function of conductivity of both p-type (square) and n-type (circle) samples. The dash line is a simulated MIM signal. (i) Average dC/dV signal as a function of conductivity. All scale bars are 2 μm.

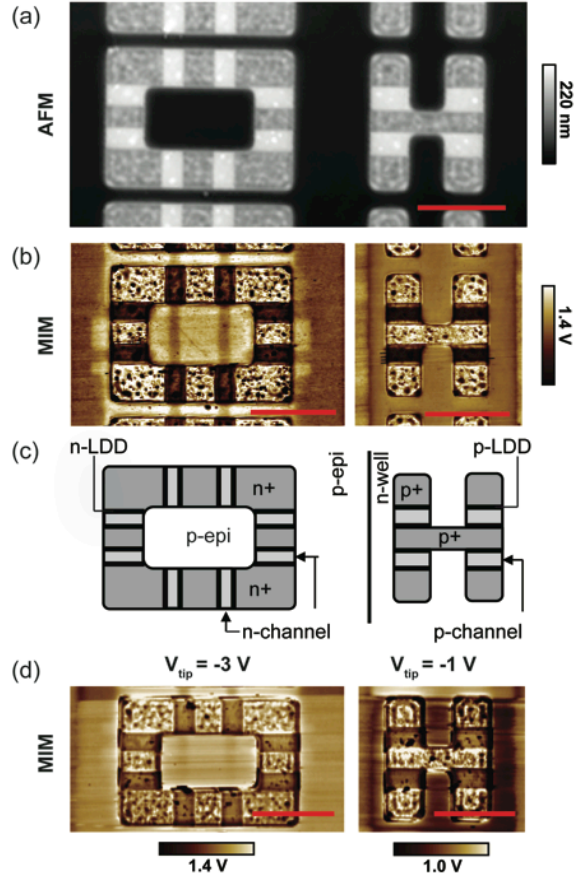

Figure 2. (color online) (a) Topography image of the SRAM sample. (b) MIM images of the rectangular and H-shape devices (c) Cartoon of the device structures, indicating n+ and p+ (dark grey), n-channel and p-channel (light grey), n-LDD and p-LDD (black), p-epi and n-well (white) regions. (d) MIM images of both devices with negative bias voltages at the tip ($V_{tip}$). All scale bars correspond to 5 μm.

| Rectangular Device | | | H-shape Device | | |
|---|---|---|---|---|---|
| | Doping type | Con. (cm$^{-3}$) | | Doping type | Con. (cm$^{-3}$) |
| p-epi | p | $2 \times 10^{16}$ | n-well | n | $2 \times 10^{17}$ |
| n-channel | p | $2 \times 10^{17}$ | p-channel | n | $1 \times 10^{17}$ |
| n-LDD | n | $5 \times 10^{18}$ | p-LDD | p | $3 \times 10^{18}$ |
| n+ | n | $2 \times 10^{20}$ | p+ | p | $4 \times 10^{19}$ |

Table 1. Nominal doping concentration of each region and the carrier type.

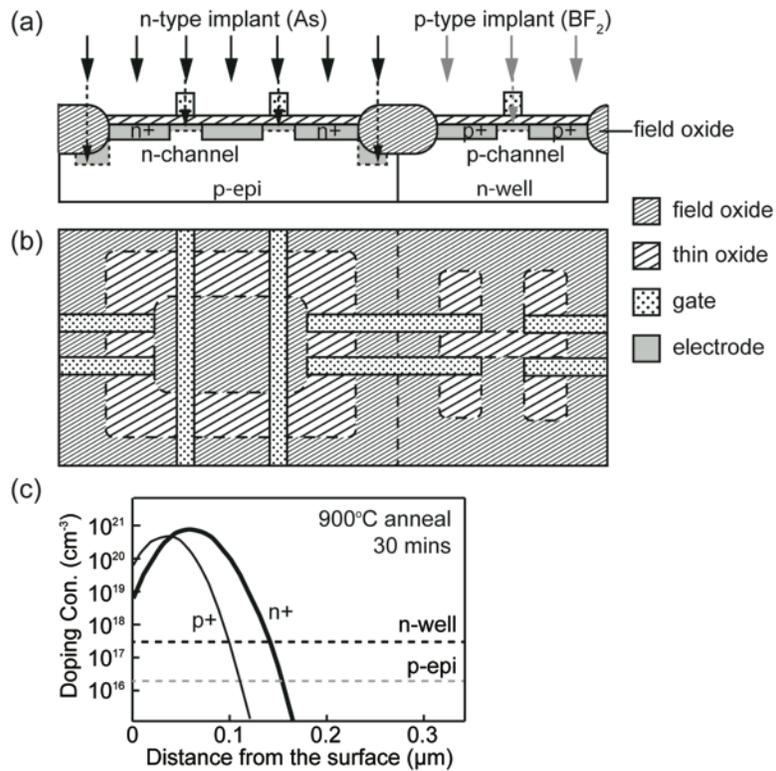

Figure 3: (a) Cross section and (b) top views of the device structure during the ion-implantation process. The thin surface implanted layers in (a) are sketched by the dotted lines. The n-type (black arrow) and the p-type (gray arrow) implants are two different steps, with photoresist covering the other device during the implantation. (c) Simulated doping profiles for the n+ (thick solid line) and p+ (thin solid line) implantation processes on $SiO_2$. The dash lines indicate the doping levels of the p-epi (gray) and the n-well (black).

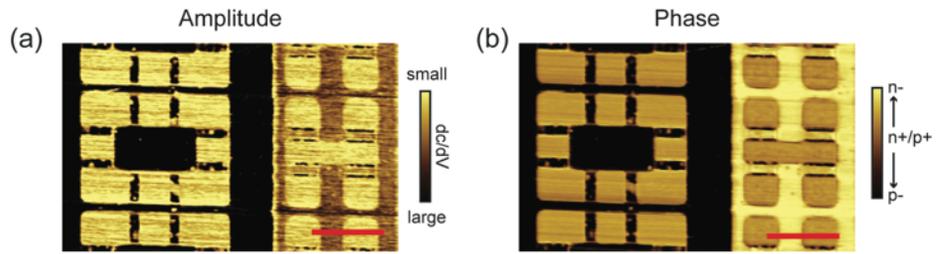

Figure 4. (color online) (a) dC/dV amplitude and (b) phase images taken by our setup. Note that the color scale in (a) is not monotonic as a function of local conductivity. All scale bars equal to 5 μm.